# Micropolis Interdependency Modeling using Open Hybrid Automata

Constantinos Heracleous

KIOS Research Center for Intelligent Systems and Networks,
Department of Electrical and Computer Engineering, University of Cyprus
heracleous.constantinos@ucy.ac.cy
http://www.kios.ucy.ac.cy/

**Abstract.** Micropolis is a virtual city that is used for various studies, such as modeling and analyzing water networks. In this paper we model the various interdependencies between three major infrastructures in Micropolis, the power system, the communication network and the water network, in order to study cascades effects between them. Specifically, we develop open hybrid automata models for the main components of the three infrastructures and then composed them together based on their interdependencies.

**Keywords:** interdependency modeling, hybrid systems, open hybrid automata

## 1 Introduction

The normal operation of the three Micropolis infrastructure systems, i.e., water, power and communication, heavily depends on the various interdependencies that exist among them. For instance, the water and communication systems receive electricity from the power system to feed the water pumps and the cell antennas, respectively, while the power and water systems rely on the communication system for data transmission for their monitoring and control. If interdependencies break due to failure in one system, there will be a disruption to the normal operation of the other systems as well. Thus, it is important to study interdependencies and develop models to analyze their cascading effects.

To facilitate the identification, understanding, and analysis of interdependencies, we usually classify them based on their characteristics into the following four principal types [1]: (a) physical, if the operations of one infrastructure depends on the physical output(s) of the other and vice versa, (b) cyber, if there is information/signal transmission between different infrastructures, (c) geographic, if components of different infrastructures are in close spatial proximity, and (d) logical, due to any other mechanism (e.g., various policy, legal, or regulatory regimes) that can link logically two or more infrastructures. Interdependencies are bidirectional relationships while dependencies are unidirectional relationships. When multiple infrastructures are connected as "system of systems" and their individual components are considered, interdependencies are the result of multiple dependencies between the components of *different* infrastructures. These are often called *external* dependencies and they can be the same type as interdependencies i.e., physical, cyber, geographic, or logical, but just unidirectional. There are also *internal* dependencies, which are the connections be-tween components inside the *same* infrastructure, for instance the connection inside the water infrastructure between a pump and a tank is an internal dependency.

## 2     Micropolis main Components and Interdependencies

For the city of Micropolis the main components of the three systems and their internal and external dependencies are shown in Figure . Specifically, the following components are considered: for the power system (1) the substation that supplies power to Micropolis and (2) the SCADA that remotely monitors and controls the substation; for the communication system (3) the network that consists of a number of microcell antennas; and lastly for the water system (4) the tank that supplies water to Micropolis and (5) the pump to which the tank is dependent for water, thus there is an internal dependency between the two. As external dependencies among the components three types are considered: (i) the physical dependencies of the pump and the network on the substation for power supply; (ii) the cyber dependencies of all the components on the network, i.e., the communication links to the substation, the SCADA, the pump and the tank; and (iii) the logical dependencies such as the power and water demands for the substation and the tank, respectively. Although there are geographic interdependencies between all the components, due to the close spatial proximity between them (Micropolis is a small town), we do not consider them in this work since they usually have an effect during intentional attacks or natural disasters (e.g., explosions, floods, and earthquakes) and given the size of Micropolis such events would affect all the components at once without any cascading effects between them.

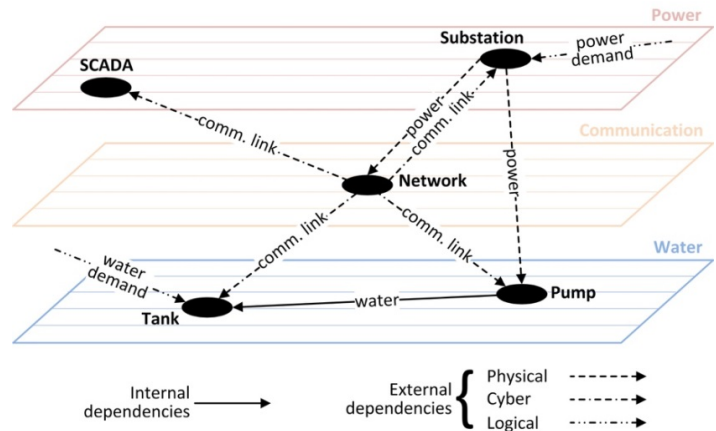

*Figure 1 Overview of Micropolis main infrastructures components and their dependencies.*

## 3     Open Hybrid Automata Modeling

To model the main components of Micropolis and the dependencies between them in order to study possible cascading effects we use open hybrid automata as proposed in [2] which the reader is encouraged to review for more information. The open hybrid automata are used as models for hybrid systems, i.e., systems with both continuous and discrete behaviors such as critical infrastructures. For instance in a water system the start and stop of a pump is discrete while the flow of water is continuous. With open hybrid automata it is possible to model the behavior of each critical infrastructure component at the necessary level of abstraction, i.e., how the component will behave in case it fails or if there is no power supply or communication available. The dependencies among the components on the other hand, are represented by the connections between the inputs and the outputs of the various open hybrid automata models. Formally, connecting the inputs and the outputs of two or more open hybrid automata is called composition, and the result is another open hybrid automaton bigger and more complex that has the remaining inputs and outputs. The new open hybrid automaton is often referred to as the composition model and consists of all the open hybrid automata running in parallel. The composition model is the one that can be simulated for various scenarios where component(s) set to fail at specific time(s) and observe the cascading effects to the other components.

*Table 1 Models variables descriptions*

| Substation | | | |
|---|---|---|---|
| $x_{t_p}$ | Timer (state) | $p_m$ | Power Measurement (output) |
| $s_{CB}$ | Switch Signal from SCADA (input) | $P_{lim}$ | Power Limit (constant) |
| $p_d$ | Power Demand (input) | $T_s$ | Safety Period (constant) |
| $p_s$ | Power Supply (output) | $\varepsilon$ | White Noise (random variable) |
| SCADA | | | |
| $x_{t_s}$ | Timer (state) | $s_{CB}$ | Switch Control Signal (output) |
| $p_m$ | Substation Power Measur. (input) | $T_d$ | Transmission Delay Time (constant) |
| $s_{OP}$ | SCADA Operator Control (input) | $T_s$ | Safety Time (constant) |
| Network | | | |
| $x_{t_{ups}}$ | Timer for UPS reserve (state) | $\phi_n$ | Fault Trigger (Input) |
| $s_{k=[1,2,3]}$ | Signals values of Substation, SCADA, and Tank, that pass and delayed through the network (input and output) | $p_{nd}$ | Instant Network Power Demand (output) |
| | | $T_{k=[1,2,3]}$ | Transmission Delay for each $s_{k=[1,2,3]}$ (constant) |
| $p_{ns}$ | Instant Power Supply (input) | $P_n$ | Network Working Power (constant) |
| Tank | | | |
| $x_v$ | Water Volume (state) | $v_{tank}$ | Instant Tank Water Volume (output) |
| $w_d$ | Water Demand (input) | $V_0$ | Initial Tank Water Volume (constant) |
| $w_s$ | Water Supply (input) | $V_{max}$ | Tank Maximum Volume (constant) |
| Pump | | | |
| $x_t$ | Timer for Pump Operation (state) | $V_{th}$ | Staring Pump Volume Threshold (constant) |
| $v_{tank}$ | Tank Water Volume Measurm. (input) | | |
| $p_{ps}$ | Instant Power Supply (input) | $T_{off}$ | Pump Min Resting Period (constant) |
| $\phi_p$ | Fault Trigger (Input) | $T_{on}$ | Pump Max Working Period (constant) |
| $w_s$ | Water Supply (output) | $W_{avg}$ | Average Water Supply (constant) |
| $p_{pd}$ | Instant Power Demand (output) | $P_p$ | Pump Working Power (constant) |
| $V_{max}$ | Tank Maximum Volume (constant) | | |

In the next subsections the open hybrid automata models for the main components of Micropolis are presented and then the composition model is derived based on the dependencies among them (see Figure ). The open hybrid automata models for the power substation, SCADA, are similar with those presented in [2], with minor modifications mostly to the variables names to make them less generic and more relevant to each infrastructure component. Descriptions for each model variables are summarized in Table 1.

## 3.1 Power Distribution Network (PDN) Components - Substation and SCADA

The open hybrid automaton model for the Power Substation is shown in Figure and is the same as the one in [2]. The model consists of two discrete states that represent the behavior of the substation i.e., Supply Power or Switch Off. When in Supply Power state the output $p_s = p_d$ (power supply to Micropolis equals to the power demand) and when in the Switch Off state the output $p_s = 0$, also $p_m$, that denotes the power measurements, changes accordingly. The transitions between the two discrete are represented by the guards. For instance, the transition from the Supply Power to the Switch Off state will occur either if its requested remotely from the SCADA ($s_{CB} = 1$) or due to safety reasons when the power demand is larger than the limit of the substation ($p_d \geq P_{lim}$). More detailed explanation for the model can be found in [2].

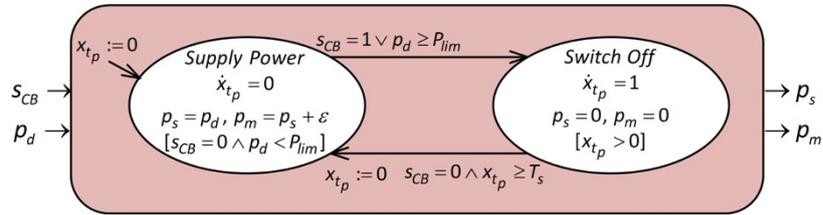

Figure 2 Power substation open hybrid automaton model.

The open hybrid automaton model for the SCADA is shown in Figure . The model is also the same as the one in [2] and represents the SCADA behavior according to the power measurements input ($p_m$) and the operator input ($s_{OP}$). The SCADA model will be at the Closed state if the substation is in the Supply Power state, representing the state of the switches at the substation, or at the Open state if the substation is in the Switch Off state or the operator decides to remotely cut the power for maintenance. In case of fault at the network it will go to the Conn. Down state. The discrete transitions are determined by the two inputs ($p_m$, $s_{OP}$) and also by the continuous state $x_{t_s}$ that acts as a timer while in the Close and Open states.

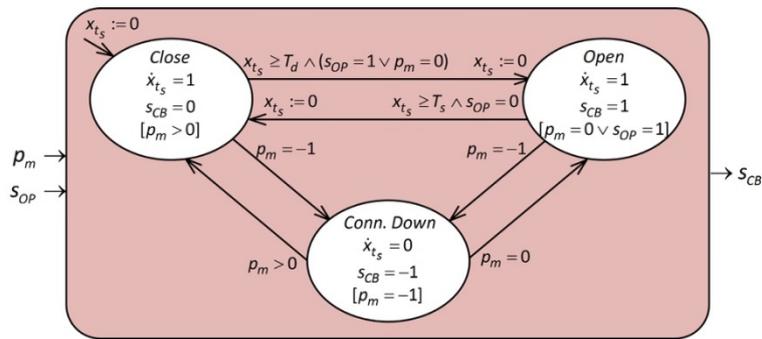

Figure 3 Power SCADA open hybrid automaton model.

## 3.2 Communication Network (CN) Component

The open hybrid automaton model for the Network is shown in Figure . The model is an extension of the work [2] with the addition of an extra signal input ($s_3$) that represents the communication between the water tank and the pump. The model consists of three discrete states that denote the behavior of the network. In the Healthy state the network operates normally providing communication services to Micropolis. In case of a power cut, where the power is not enough to feed the antennas ($p_{ns} < P_n$) they will switch to UPS, thus the model transitions to the UPS usage state. However, the UPS can last for only a certain amount of time and this is what the continuous state $x_{t_{ups}}$ is counting. Once it reaches the limit ($x_{t_{ups}} \geq T_{ups}$) the network will stop operating and this is represented in the model by transitioning to the Net Down state. Transition to the Net Down state is also triggered in case of a technical fault in the network which is denoted with the input $\phi_n$. More detailed explanation for the model can be found in [2].

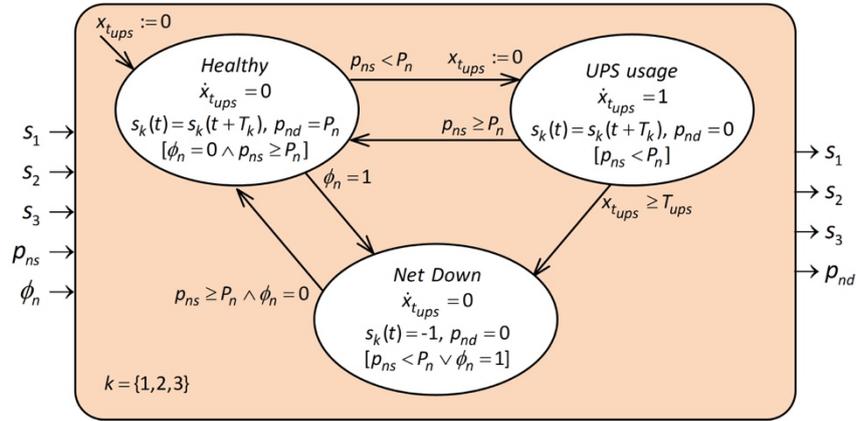

Figure 4 Network open hybrid automaton model

## 3.3 Water Distribution Network (WDN) Components - Tank and Pump

The tank receives and stores water from the pump and supplies Micropolis based on the demand. This has to be controlled accordingly since too much water from the pump can overflow the tank while less water can drain it. The tank open hybrid automaton model is shown in Figure . The continuous state $x_v$ denotes the volume of the tank that changes according to $\dot{x}_v = w_s - w_d$, where $w_s$ denotes the water supply rate from the pump, and $w_d$ the water demand rate of consumers. The discrete states of the model are determined by the tank volume. For instance, the model will be at the "Healthy" state while $0 < x_v \leq V_{max}$, where $V_{max}$ is the tank's maximum volume, and it will transition to either the Drained state if $x_v \leq 0$ or to the Overflow state if $x_v > V_{max}$. The model will return back to the "Healthy" state if the water supply becomes larger than the demand $w_s > w_d$ and is in the Drained state, or if the water demand becomes larger than the supply $w_d > w_s$ and is in the Overflow state. The single output of the model $v_{tank}$ denotes the tank's volume measurement depending on the state, and is transmitted to the pump through the network so that the pump can start and stop as described next.

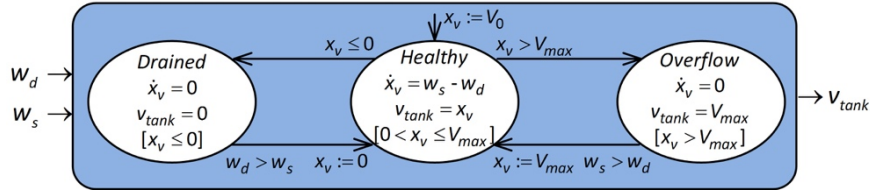

*Figure 5 Tank open hybrid automaton model.*

The pump supplies Micropolis tank with water by receiving and comparing the tank volume measurement with some threshold, if the tank volume goes below the threshold value then the pump will start given that the substation provides the necessary power. There are also restrictions to the pump operation such as maximum working period and minimum resting period. The pump open hybrid automaton model is shown in Figure . The model has three discrete states: (i) the *Pump Off* when the pump is off, (ii) the *Pump On* when the pump is on and supplies the tank with water, and (iii) the *Fault state* when there is power cut, network issues, or technical fault that prevent the pump to operate normally. The discrete transitions between the *Pump Off* and *Pump On* states are trigger by the input $v_{tank}$, that denotes the tank's volume measurement, and by the continuous state $x_t$, that acts as a timer counting the working and the resting period of the pump. Specifically, the transition from *Pump Off* to *Pump On* is triggered when the tank's volume goes below the specified threshold $V_{th}$ and also the pump's resting period $T_{off}$ has expired, i.e. $v_{tank} < V_{th} \wedge x_t \geq T_{off}$. The transition from the *Pump On* back to *Pump Off* state is triggered either when the tank's water volume reaches the maximum volume $V_{max}$ or if the maximum working period $T_{on}$ of the pump has been reached, i.e. $v_{tank} \geq V_{max} \vee x_t > T_{on}$. In case of one or more issues such as: less power supply than what is necessary ($p_{ps} < P_p$), network issues ($v_{tank} = -1$), or the pump suffers a technical fault ($\phi_p = 1$), the model will transition to the Fault state. Once the issue(s) is/are resolved it will transition to the *Pump Off* state and it will be ready to supply water to the tank accordingly. The two outputs of the model, $w_s$ that denotes the water supply by the pump and $p_{pd}$ that denotes the power demand of the pump, change values based on the discrete state of the model. In both *Pump Off* and *Fault* states $w_s = 0$ and $p_{pd} = 0$, while in the *Pump On* $w_s = W_{avg}$ and $p_{pd} = P_p$, where $W_{avg}$ is the average water supply rate of the pump and $P_p$ the necessary pump power.

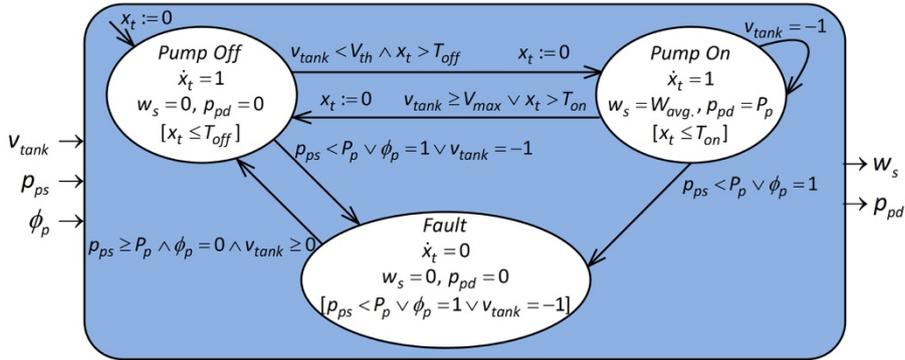

*Figure 6 Pump open hybrid automaton model.*

## 3.4 Composition Model

The five open hybrid automata models are composed together as shown in Figure creating the composition model, a larger model that includes the various dependencies between the components. As elaborated above, there are physical, cyber and logical external dependencies between the components and internal dependency between the tank and the pump, that are all depicted in detail in Figure . In the composition the component models run in parallel with the output of one model to become the input to the other, developing various feedback loops between them, which represent infrastructure interdependencies. For example, both the power substation and the SCADA use the network to transmit power measurements and control signals between them, as depicted in Figure with the necessary connections. The network is also depended on the substation model for power. These dependency connections create feedback loops between these three models, which subsequently form an interdependency between the power and the communication infrastructure. This allows the study of cascading effects between the various components by running simulation scenarios with the composition model as presented in the next section.

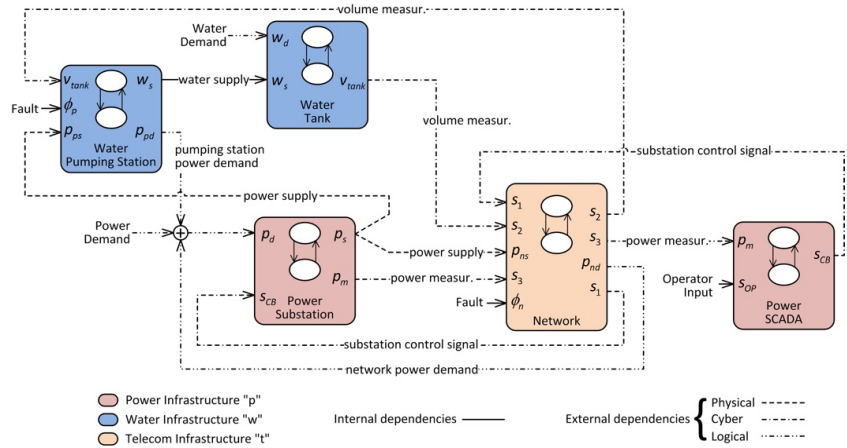

*Figure 7 Composition model*

## 4 Conclusions

In this work we have develop open hybrid automata models of the main critical infrastructures for Micropolis virtual city, that can be used to study cascading faults. Specifically, the composition model, through several simulation scenarios, can determine under what conditions a fault in one component can cascade to other components in other infrastructures due to the effects of interdependencies.